\documentclass[a4paper]{jpconf}

\usepackage{amssymb}

\usepackage{amsmath}
\usepackage{amsmath,amssymb,amsthm}
\usepackage{multicol}
\usepackage{color}
\usepackage{graphicx}
\usepackage{hyperref}

\theoremstyle{definition}

\newcommand{\gf}{\mathfrak{g}}
\newcommand{\af}{\mathfrak{a}}

\newcommand{\hf}{\mathfrak{h}}

\newcommand{\gfh}{\hat{\mathfrak{g}}}
\newcommand{\afh}{\hat{\mathfrak{a}}}

\theoremstyle{definition}

\begin{document}

\title{Algebraic properties of CFT coset construction and Schramm-Loewner evolution}
\author{A A Nazarov$^{1,2}$}
\address{
  $^1$ Department of High-Energy and Elementary Particle Physics, 
  Faculty of Physics, \\ SPb State University
  198904, Saint-Petersburg, Russia}
\address{
  $^{2}$ Chebyshev Laboratory,
  Faculty of Mathematics and Mechanics, \\ SPb State University
  199178, Saint-Petersburg, Russia}

\ead{antonnaz@gmail.com}

\begin{abstract}
  Schramm-Loewner evolution appears as the scaling limit of interfaces in lattice models at critical point. Critical behavior of these models can be described by minimal models of conformal field theory. Certain CFT correlation functions are martingales with respect to SLE. 
  We generalize Schramm-Loewner evolution with additional Brownian motion on Lie group $G$ to the case of factor space $G/A$. We then study connection between SLE description of critical behavior with coset models of conformal field theory. In order to be consistent such construction should give minimal models for certain choice of groups.

\end{abstract}

\section{Introduction}
Schramm-Loewner evolution (SLE) was introduced by Oded Schramm  \cite{schramm2000scaling} to describe scaling limit of critical interfaces in two-dimensional statistical lattice models. This approach to the critical systems has lead to numerous strict results in the study of critical behavior (see reviews  \cite{rohde2005basic}, \cite{bauer20062d}, \cite{Cardy:2005kh}). We give some basics of SLE in Section \ref{sec:schr-loewn-evol}.  

Probability measure induced by Schramm-Loewner evolution on random curves is conformally invariant. Since conformal field theory (CFT) gives us another tool-set for the study of two-dimensional critical systems, it is natural to discuss its connection with SLE.  This connection  was studied in \cite{bauer2004conformal,bauer2004cfts,bauer2003sle,bauer2002sle} and many other papers mostly for unitary minimal models.
The idea is to consider certain observables in the domain with the cut. This construction is briefly discussed in Section \ref{sec:corr-betw-sle}. 

More general models of CFT has additional symmetries. For example Kac-Moody algebras  appear in Wess-Zumino-Novikov-Witten models. To introduce such symmetry in SLE approach one considers additional random motion on Lie group \cite{bettelheim2005stochastic}, \cite{Rasmussen:2004xr}. This construction is described in Section \ref{sec:sle-wzw-models}. The correspondence between SLE martingales in this model and WZNW-correlation functions was studied in \cite{alekseev2010sle}. Similar generalization to $\mathbb{Z}(N)$-parafermionic models was proposed in \cite{santachiara2008sle,picco2008numerical}. 

Main task of present paper is to generalize SLE with additional random walk on Lie group to the case of factor space $G/H$ and to study the connection with coset construction  of conformal field theory. We remind some details of coset construction, introduce SLE on factor space and relate it to conformal field theory in Section  \ref{sec:coset-models}. Similarly to WZNW-case we obtain the system of algebraic equations on boundary condition changing operators from martingale condition. 
Coset construction of CFT can be used to obtain minimal models and parafermions (see \cite{difrancesco1997cft}) so we compare our equations on boundary condition changing operators with the results of paper \cite{santachiara2008sle}.

In Conclusion \ref{sec:conclusion} we discuss the need to compare classification of boundary condition changing operators which follows from martingale condition with general classification of b.c.c. operators in boundary conformal field theory related to D-brane solutions \cite{fuchs2005geometry,fredenhagen2002d,elitzur2002d,Maldacena:2001ky,felder1999geometry,alekseev1999d}. 

\subsection{Schramm-Loewner evolution}
\label{sec:schr-loewn-evol}
\begin{figure}[h]
  \centering{
    \includegraphics[height=50mm]{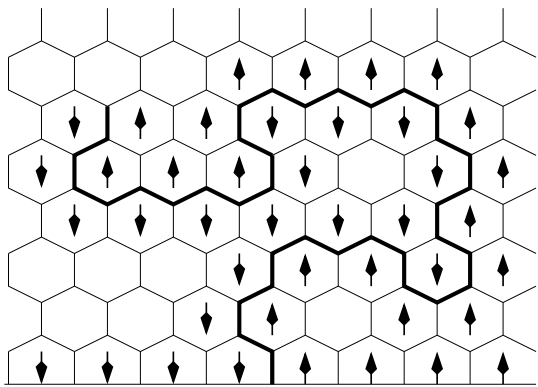}
    \caption{Interface in Ising model on triangular lattice}}
  \label{fig:sle}
\end{figure}

Consider Ising model on triangular lattice on upper half plane (see Fig. \ref{fig:sle}). We impose following boundary condition: all spins are down on one half of the boundary and all spins are up on another half. Then in any spin configuration we get an interface delimiting two clusters and connecting zero and infinity (see Figure \ref{fig:sle}).

We can impose the condition of the existence of finite-length part of interface and study statistical model with this condition. Obviously it is equivalent to the model in the slit domain with the cut along an interface. 

\begin{figure}[h]
  \centering{
    \includegraphics[width=60mm]{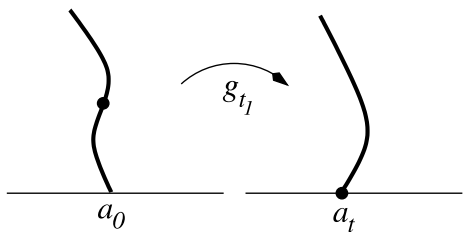}
    \caption{Conformal map $g_{t}(z):\mathbb{H}_{t}\to \mathbb{H}$.}}
  \label{fig:sle2}
\end{figure}

Now consider continuous limit of lattice model with a (part of) interface of finite length. Interface in random configuration of the model tends to random curve. Old hypothesis \cite{Polyakov:1970xd} of conformal invariance at critical point was recently strictly proved for some lattice models (See \cite{smirnov2007towards}, \cite{duminil2011conformal} for a review). We assume conformal invariance at critical point and consider upper half-plane $\mathbb{H}$ with the cut along a critical interface $\gamma_{t}$. We denote this slit domain by $\mathbb{H}_{t}=\mathbb{H}\setminus \gamma_{t}$. Conformal map from $\mathbb{H}_{t}$ to $\mathbb{H}$ is denoted by $g_{t}:\mathbb{H}_{t}\to \mathbb{H}$ (See Figure \ref{fig:sle2}).

In paper \cite{schramm2000scaling} it was shown that $g_{t}(z)$ satisfies stochastic differential equation 
\begin{equation}
\label{eq:19}
  \frac{\partial g_t(z)}{\partial t} = \frac{ 2}{g_t(z)-\sqrt{\kappa}\xi_{t}} ,
\end{equation}
where $\xi_{t}$ is the Brownian motion. The dynamic of the tip $z_{t}$ of critical curve $\gamma_{t}$ (tip of SLE trace) is given by the law $z_{t}=g_{t}^{-1}(\sqrt{\kappa}\xi_{t})$. 

Stochastic process which satisfies \eqref{eq:19} is called     {\it Schramm-Loewner evolution} on the upper half-plane $\mathbb{H}$.
For us it is more convenient to use map $w_{t} (z)=g_{t}(z)-\sqrt{\kappa}\xi_{t}$, so the equation \eqref{eq:19} becomes
\begin{equation}
  \label{eq:20}
       d w _{t}= \frac{2dt}{w_{t} }-\sqrt{\kappa}\xi_{t}  
\end{equation}
Schramm-Loewner evolution provides conformally-invariant probability measure on curves $\gamma_{t}$ in $\mathbb{H}$.

\subsection{SLE martingales and CFT correlation functions}
\label{sec:corr-betw-sle}

Now we can look at the observables in the presence of SLE trace. Expectation value of lattice observable $\mathcal{O}$ on the upper half-plane can be calculated as the sum of expectation values of this observable in presence of SLE trace $\gamma_{t}$ up to some time $t$ multiplied by the probability of this trajectory:
\begin{equation*}
  \prec \mathcal{O} \succ_{\mathbb{H}}=\mathbb{E}\left[\prec\mathcal{O}\succ_{\gamma_{t}}\right]=\sum_{\gamma_{t}} P\left[C_{\gamma_{t}}\right] \prec \mathcal{O} \succ_{\gamma_{t}}
\end{equation*}
Lattice observable  $\prec \mathcal{O} \succ_{\mathbb{H}}$ does not depend on $t$, hence $\prec\mathcal{O}\succ_{\gamma_{t}}$ is a martingale.

In continuous limit lattice observable tends to CFT correlation function \cite{bauer2003sle,bauer2003conformal,bauer2002sle}.
We need to take into account the change of boundary conditions on the tip of SLE trace and at the infinity, so the expression is
\begin{equation}
  \prec \mathcal{O} \succ_{\mathbb{H}_{t}}\to \mathcal{F}(\left\{z_{i}\right\})_{\mathbb{H}_{t}}=
  \frac{\left< \mathcal{O}(\{z_{i}\})\phi(z_{t})\phi^{\dagger}(\infty)\right>_{\mathbb{H}_{t}}}{\left<\phi(z_{t})\phi^{\dagger}(\infty)\right>_{\mathbb{H}_{t}}}
\label{eq:21}
\end{equation}


We consider the theory with the boundary, so in principle we need to use boundary conformal field theory here and impose proper consistency conditions \cite{cardy1984conformal,cardy1989boundary,cardy1991bulk}. In the case of upper half-plane correlation functions of boundary CFT can be rewritten as the correlation functions of the theory on the whole plane but with doubled number of bulk fields.


We assume that $\mathcal{F}$ contains some set of primary fields $\varphi_{\lambda_{i}}$ with conformal weights $h_{\lambda_{i}}$. Since we work with boundary CFT we need to add bulk fields in conjugate points $\bar z_{i}$. For future use in WZNW models we denote the conformal weights of these duplicate fields by $h_{\lambda_{i}^{*}}$.  Also we have boundary condition changing operators  $\phi$ at the tip of SLE trace and at the infinity.  We use the conformal map  $w(z):\mathbb{H}\setminus\gamma_{t}\to \mathbb{H}$ to rewrite expression \eqref{eq:21} in the whole upper half plane:

\begin{equation}
  \mathcal{F}(\left\{z_{i}\right\})_{\mathbb{H}_{t}}=\prod \left(\frac{\partial w(z_{i})}{\partial z_{i}}\right)^{h_{\lambda_i}} 
  \prod \left(\frac{\partial \bar w(\bar z_{i})}{\partial \bar z_{i}}\right)^{h_{\lambda^{*}_i}}
  \mathcal{F}(\left\{w_{i}, \bar w_{i}\right\})_{\mathbb{H}}
  \label{eq:1}
\end{equation}

Now we need to consider the evolution of SLE trace $\gamma_{t}$ from  $t$ to $t+ dt$. First factor in right-hand-side of equation (\ref{eq:1}) gives us
\begin{equation*}
  -\frac{2h_{\lambda_{i}}}{w_{i}^{2}}\left(\frac{\partial w_{i}}{\partial z_{i}}\right)^{h_{\lambda_{i}}}.
\end{equation*}
For transformation of primary fields $\varphi_{\lambda_{i}}$ we have 
\begin{equation}
  \label{eq:2}
  d\varphi_{\lambda_{i}}(w_{i}) = \mathcal{G}_{i}\varphi_{\lambda_{i}}(w_{i})=\left(\frac{2dt}{w_{i}}-\sqrt{\kappa} d\xi_{t}\right) \partial_{w_{i}}\varphi_{\lambda_{i}}(w_{i}) 
\end{equation}
We denote generator of this transform by $\mathcal{G}_{i}$.

Since $ \prec\mathcal{O}\succ_{\gamma_{t}}$ is a martingale, the expectation value of its increment during the evolution from $t$ to $t+dt$ is zero
\begin{equation}
  \mathbb{E}\left[\prec\mathcal{O}\succ_{\gamma_{t}}\right]=    \mathbb{E}\left[\prec\mathcal{O}\succ_{\gamma_{t+dt}}\right], \quad d\mathbb{E}\left[ \prec\mathcal{O}\succ_{\gamma_{t}}\right]=0
\label{eq:22}
\end{equation}
Using Ito calculus we get the expression for the differential of $\mathcal{F}$:
\begin{equation}
d \mathcal{F}_{\mathbb{H}_{t}}= \left(\prod_{i=1}^{2N}\left(\frac{\partial w_{i}}{\partial z_{i}}\right)^{h_{i}}\right)\left(-\sum_{i=1}^{2N}\frac{2h_{i}dt}{w_{i}^{2}}+\left[\sum_{i=1}^{2N}\mathcal{G}_{i}+\frac{1}{2}
      \sum_{i,j}\mathcal{G}_{i}\mathcal{G}_{j}\right]\right)\mathcal{F}_{\mathbb{H}}=0
\label{eq:8}
\end{equation}
We substitute formula \eqref{eq:2} and obtain the equation
\begin{equation}
  \left( \sum_{i}\left[-\frac{2h_{i}}{w_{i}^{2}} +\frac{2}{w_{i}}\partial_{w_{i}}\right]+\frac{\kappa}{2}\sum_{i,j}\partial_{w_{i}} \partial_{w_{j}}\right)\mathcal{F}(\left\{z_{i}\right\})=0
\label{eq:23}
\end{equation}

For the correlation functions of secondary fields $L_{-n}\phi$ we have the equation 
\begin{equation}
\left< (L_{-n}\phi)(z) \varphi_{1}\dots \varphi_{N}\right>=\mathcal{L}_{-n}\left<\phi\varphi_{1}\dots\varphi_{N}\right>,
\end{equation}
 where $n\geq 1$ and
\begin{equation*}
  \mathcal{L}_{-n}=\sum_{i=1}^{N} \left(\frac{(n-1)h_{i}}{(w_{i}-z)^{n}} -\frac{1}{(w_{i}-z)^{n-1}}\partial_{w_{i}}\right)
\end{equation*}
(See \cite{difrancesco1997cft}). So we can rewrite the differential equation \eqref{eq:23} as the algebraic equation on the field $\phi$ corresponding to boundary condition changing operator:
\begin{equation}
  \label{eq:24}
   \left<\left([L_{-2}-\frac{\kappa}{2}L_{-1}^{2}]\phi\right)(0)\; \varphi_{\lambda_{1}}\dots \varphi_{\lambda_{{2N}}}\right>=0.
\end{equation}
In the case of minimal models the set of primary fields is finite and all the states are obtained through field-state correspondence \cite{belavin1984ics}, \cite{difrancesco1997cft}. 
Equation \eqref{eq:24}  holds for arbitrary observable $\mathcal{O}$ and arbitrary primary fields $\varphi_{\lambda_{i}}$, so $\psi=[L_{-2}-\frac{\kappa}{2}L_{-1}^{2}]\phi$ is level two null field and $\psi(0)\left|0\right>$ is level two null-state. There exist only two primary fields in minimal unitary models which give rise to Verma modules of Virasoro algebra with level two null-states. They are $\phi_{1,2}$ and $\phi_{2,1}$, so $\phi\sim \phi_{1,2} \;\text{or}\; \phi_{2,1}$. 

\section{Wess-Zumino-Novikov-Witten models}
\label{sec:sle-wzw-models}
To generalize the analysis of previous section \ref{sec:corr-betw-sle} to non-minimal models we need to take into account extra symmetries. Wess-Zumino-Novikov-Witten models has Kac-Moody symmetry in addition to conformal invariance leading to the appearance of Virasoro algebra. At first we remind well-known properties of WZNW-models \cite{difrancesco1997cft} which are useful to study SLE martingales. 

 The action of WZNW model can be written in terms of map $g:\mathbb{C}\cup \{\infty\}\sim S^{2}\to G$ from complex plane with infinity or two-sphere to some (simple) Lie group $G$:
\begin{multline}
  S=-\frac{k}{8\pi}\int d^2x\; \mathcal{K} (g^{-1}\partial^{\mu}g, g^{-1} \partial_{\mu}g)  
  - \frac{k }{24\pi^{2}} \int_{B}\epsilon_{ijk} \mathcal{K}\left(
    \tilde g^{-1}\frac{\partial \tilde g}{\partial y^i},\left[
      \tilde g^{-1}\frac{\partial \tilde g}{\partial y^j}
      \tilde g^{-1}\frac{\partial \tilde g}{\partial y^k}\right]\right) d^3y
\end{multline}
 Here $\mathcal{K}$ is Killing form on Lie algebra $\gf$ of Lie group $G$. The first term is just non-linear sigma-model and the second term is written using the continuation $\tilde{g}$ from two-sphere $S^{2}$ to three-dimensional manifold $B$ which has two-sphere as the boundary. Since this continuation is non-unique we get the requirement for $k$ to be integer. 

Conserved currents of this model have the following form:
  \begin{eqnarray}
    J(z)= -k \partial_zg g^{-1}
    \bar J(\bar z)=k g^{-1}\partial_{\bar z}g
  \end{eqnarray}

The model possesses gauge invariance under the transformations
  $g(z,\bar z)\to \Omega(z)g(z,\bar z)\bar \Omega^{-1}(\bar z)$, 
where $\Omega,\;\bar \Omega \in G$ are are independent. If we consider infinitesimal gauge transformation $\Omega=1+\omega$ we get Ward identities 
  \begin{equation}
    \label{eq:87}
    \delta_{\omega,\bar \omega}\left< X \right>=-\frac{1}{2\pi i}\oint dz \sum\omega^a \left< J^a X\right>+
    \frac{1}{2\pi i} \oint d\bar z \sum \bar \omega^a \left< \bar J^a X\right>,
  \end{equation}
which can be used to get operator product expansion for currents:
 \begin{equation}
   \label{eq:3}
   J^{a}(z)J^{b}(w)\sim \frac{k\delta_{ab}}{(z-w)^{2}}+\sum_{c}i f_{abc}\frac{J^{c}}{z-w},
 \end{equation}
where $f_{abc}$ are the structure constants of Lie algebra $\gf$.
 
Expanding the currents to modes
$J^a(z)=\sum\limits_{n\in \mathbb Z}z^{n-1}J^a_n $
and using operator product expansion \eqref{eq:3} we get commutation relations of affine Lie algebra $\gfh$:
\begin{equation*}
 \left[J^a_n,J^b_m\right]=\sum_c i f^{abc}J^c_{n+m}+kn\delta^{ab}\delta_{n+m,0}.  
\end{equation*}

Conformal invariance of the model can  be seen from 
Sugawara construction, which is a way to embed Virasoro algebra into the universal enveloping algebra of affine Lie algebra $\gfh$ ($Vir\subset U(\gfh)$):
\begin{equation}
  \label{eq:4}
  L_n=\frac{1}{2(k+h^{\vee})}\left(\sum\limits_a\sum\limits_{m\leq -1}J^a_m J^a_{n-m}+\sum_{m\geq 0} J^{a}_{n-m}J^{a}_{m}\right), 
\end{equation}
 where $h^{\vee}$ is the dual Coxeter number of Lie algebra $\gf$. 
Full chiral algebra of the model is semidirect product of affine and Virasoro algebra $\gfh \ltimes Vir$ with commutation relations 
  \begin{equation}
    \label{eq:92}
    \begin{aligned}
      \left[J^a_n,J^b_m\right]=\sum_c i f^{abc}J^c_{n+m}+kn\delta^{ab}\delta_{n+m,0} \\
      \left[L_n,L_m\right]=(n-m)L_{n+m}+\frac{c}{12}(n^3-n)\delta_{n+m,0}\\
      \left[L_n,J^a_m\right]=-mJ^a_{n+m}
    \end{aligned}
  \end{equation}
Here central charge $c$ of Virasoro algebra is given by expression
\begin{equation}
  \label{eq:25}
  c=\frac{k\;\mathrm{dim}\gf}{k+h^{\vee}}
\end{equation}
There is infinite number of Virasoro primary fields, but they are organized into highest weight modules of affine Lie algebra $\gfh$. 
Primary fields of full chiral algebra $\phi_{\lambda}$ are labeled by highest weights of  $\gfh$-modules. Here we see how Virasoro and affine Lie algebra generators act on primary fields:
  \begin{equation*}
    \begin{aligned}
      & J_0^a\left|\phi_{\lambda}\right>=-t^a_{\lambda}\left|\phi_{\lambda}\right>  \quad    J^a_n\left|\phi_{\lambda}\right>=0 \quad \mbox{for}\; n>0 \\
      & L_0\left|\phi_{\lambda}\right>=\frac{1}{2(k+h^{\vee})}\sum_aJ^a_0J^a_0\left|\phi_{\lambda}\right>=\frac{(\lambda,\lambda+2\rho)}{2(k+h^{\vee})}\left|\phi_{\lambda}\right>=h_{\lambda} \left|\phi_{\lambda}\right>,
    \end{aligned}
  \end{equation*}
where $\rho$ is Weyl vector of $\gf$. 

Now we want to study Schramm-Loewner evolution in WZNW-models.
Similarly to minimal models we consider observable
\begin{equation*}
  \mathcal{F}(\left\{z_{i}\right\})_{\mathbb{H}_{t}}=
  \frac{\left<\phi_{\Lambda}(z_{t}) \phi_{\lambda_1}(z_{1}) \dots \phi_{\lambda_n}(z_{n}) \phi_{\lambda^{*}_1}(\bar z_{1}) \dots \phi_{\lambda^{*}_n}(\bar z_{n})
      \phi_{\Lambda^{*}}(\infty)\right>}{\left<\phi_{\Lambda}(z_{t})\phi_{\Lambda^{*}}(\infty)\right>}
\end{equation*}
Again we can use conformal map  $w(z):\mathbb{H}\setminus\gamma_{t}\to \mathbb{H}$ to rewrite it on the whole upper half-plane. 
\begin{equation*}
  \mathcal{F}(\left\{z_{i}\right\})_{\mathbb{H}_{t}}=\prod \left(\frac{\partial w(z_{i})}{\partial z_{i}}\right)^{h_{\lambda_i}} 
  \prod \left(\frac{\partial \bar w(\bar z_{i})}{\partial \bar z_{i}}\right)^{h_{\lambda^{*}_i}}
  \mathcal{F}(\left\{w_{i}, \bar w_{i}\right\})_{\mathbb{H}}
\end{equation*}

Under the evolution from $t$ to $t+dt$ first factor gives us $-\frac{2h_{\lambda_{i}}}{w_{i}^{2}}\left(\frac{\partial w_{i}}{\partial z_{i}}\right)^{h_{\lambda_{i}}}$.

When we consider fields we need to add random gauge transformation (random motion in $G$) to stochastic evolution \cite{bettelheim2005stochastic}, \cite{alekseev2010sle}. 
For fields we have
\begin{equation*}
  d\phi_{\lambda_{i}}(w_{i}) = \mathcal{G}_{i}\phi_{\lambda_{i}}(w_{i}),
\end{equation*}
so additional term appears in the generator of field transformation. 
\begin{equation}
  \mathcal{G}_{i}=\left(\frac{2dt}{w_{i}}-\sqrt{\kappa} d\xi_{t}\right) \partial_{w_{i}}+\frac{\sqrt{\tau}}{w_{i}}\sum_{a=1}^{\mathrm{dim} \gf}\left(d \theta ^{a} t^{a}_{i}\right)
\label{eq:18}
\end{equation}
Here $d\theta^{a}$ are the generators of $\mathrm{dim}\gf$-dimensional Brownian motion and $\mathbb{E}[d\theta^{a}d\theta^{b}]=\delta_{ab}dt$. We assume that $t^{a}$ is a basis of $\gf$ orthogonal with respect to Killing form $\mathcal{K}$, $\mathcal{K}(t^{a},t^{b})=\delta_{ab}$.

Using formula \eqref{eq:8} we get the equation from martingale condition:
\begin{equation}
  \left(-2 \mathcal{L}_{-2}+\frac{1}{2}\kappa \mathcal{L}_{-1}^{2}+\frac{1}{2}\tau\sum_{a} \mathcal{J}^{a}_{-1} \mathcal{J}^{a}_{-1}\right)        \mathcal{F}(\left\{w_{i}, \bar w_{i}\right\})_{\mathbb{H}}=0,
  \label{eq:27}
\end{equation}
where
\begin{equation*}
  \mathcal{L}_{-n}=\sum_{i}\left(\frac{(n-1)h_{\lambda_{i}}}{(w_{i}-z)^{n}}-\frac{1}{(w_{i}-z)^{n-1}}\partial_{w_{i}}\right);\quad \mathcal{J}^{a}_{{-n}}=-\sum_{i}\frac{t^{a}_{i}}{(w_{i}-z)^{n}}
\end{equation*}
Again we can rewrite it as algebraic requirement for  field  which corresponds to boundary condition changing operator. Now
\begin{equation}
  \left| \psi\right>=\left(-2 L_{-2}+\frac{1}{2}\kappa L_{-1}^{2}+\frac{1}{2}\tau\sum_{a} J^{a}_{-1} J^{a}_{-1}\right) \left|\phi_{\Lambda}\right>    
  \label{eq:16}
\end{equation}
is level two null state and if we act on this state with raising operators we should get zero.
\begin{eqnarray}
  J^{a}_{1} \left|\psi\right>=0\\
  J^{a}_{2}\left|\psi\right>=0\\
  L_{2}\left|\psi\right>=0\\
  L_{1}\left|\psi\right>=0
\end{eqnarray}
Using commutation relations \eqref{eq:92} these equations can be rewritten as the algebraic equations connecting parameters of random motion $\kappa, \tau$ with the level $k$ of affine Lie algebra representation. Rigorous analysis of these equations is given in the paper \cite{alekseev2010sle}.

\section{Coset models}
\label{sec:coset-models}
Now we want to generalize analysis of correspondence between SLE and CFT even further and study coset models of conformal field theory\cite{Goddard198588}. Such models are specified by  Lie algebra $\gf$ and its subalgebra $\af$. Denote by $J_{n}^{a}$ the generators of affine Lie algebra $\gfh$ and by $\tilde{J}_{n}^{b}$ the generators of $\afh$, so that $\tilde{J}^{b}_{n}=\sum_{a} m_{a}^{b} J^{a}_{n}$.
Virasoro generators in coset models are given by the difference of Sugawara expressions of $\gf$ and $\af$-WZNW models:
\begin{equation*}
  L_{n}=L_{n}^{\gf}-L_{n}^{\af}
\end{equation*}
Commutation relations of Virasoro generators with generators of subalgebra $\afh$ are trivial
\begin{equation}
  \label{eq:26}
  \left[L_{n},\tilde{J}^{b}_{m}\right]=0
\end{equation}

Coset models can be realized as gauged Wess-Zumino-Novikov-Witten models by adding gauge fields  $A, \bar{A}$ taking values in Lie algebra $\af\subset \gf$ to the action\cite{gawdzki1988g}. Then fields are labeled by pairs of weights $(\mu,\nu)$, where $\mu$ is the weight of $\gfh$ and $\nu$ is the weight of $\afh$ correspondingly. But there are selection rules, i.e. the branching functions for certain pairs of weights $(\mu,\nu)$ vanish. There is also a redundancy, i.e. non-vanishing branching functions for distinct pairs $(\mu,\nu)$ turn out to be identical \cite{fuchs1996resolution,schellekens1990field}.

So, primary fields are labeled by pairs of weights $(\mu,\nu)\in \hf_{\gfh}\oplus \hf_{\afh}$  of algebra $\gfh$ and subalgebra $\afh$, such that branching functions $b^{\mu}_{\nu}(q)\neq 0$. Some pairs are equivalent. This equivalence is given by the action of so called ``simple currents'' $(J,\tilde{J})$, which are certain elements of outer  automorphisms group $\mathcal{O}(\gfh)\times \mathcal{O}(\afh)\approx B(G)\times B(A)$ of $\gfh\times\afh$, where $B(G)$ is center of Lie group G. We can think of simple currents as of primary fields and their action on fields of theory are then given by fusion product \cite{difrancesco1997cft}.  So primary fields of coset model are given by the equivalence classes of pairs of weights $(\mu,\nu)\sim (J*\mu,\tilde{J}*\nu)$, where $(J,\tilde J)$ such that their conformal weights are equal:  $h_{J}-h_{\tilde{J}}=0$. 

Conformal weight of coset primary field is equal to
\begin{multline}
  L_0\left|\phi_{(\mu,\nu)}\right>=\left(\frac{1}{2(k+h^{\vee})}\sum_aJ^a_0J^a_0-\frac{1}{2(k+h_{\af}^{\vee})}\sum_b \tilde{J}^b_0 \tilde{J}^b_0 \right)
  \left|\phi_{(\mu,\nu)}\right>=\\
  \left(\frac{(\mu,\mu+2\rho)}{2(k+h^{\vee})}-\frac{(\nu,\nu+2\rho_{\af})}{2(k+h^{\vee})}\right)\left|\phi_{(\mu,\nu)}\right>
\end{multline}
It is possible to obtain analogues of Knizhnik-Zamolodchikov equations for coset models \cite{kogan1997knizhnik}:
\begin{equation}
  \left\{\frac{1}{2}\partial_{i} + \sum_{i\neq j}^{N}\left(\frac{t^{a}_{i}t^{a}_{j}}{k+h^{\vee}}-\frac{\tilde t^{b}_{i}\tilde t^{b}_{j}}{k+h^{\vee}_{\af}}\right)\frac{1}{z_{i}-z_{j}}\right\} \left<\phi_{1}(z_{1})\dots \phi_{N}(z_{N})\right>=0
  \label{eq:6}
\end{equation}

Let us introduce Schramm-Loewner evolution corresponding to coset models. 
The idea is to constrain additional Brownian motion on group manifold to the factor space $G/A$.

To study SLE on $G/A$ factor space we restrict random walk on group manifold by the choice of basis in $\gf$. Assume that generators $\{J^{a};\; a=1\dots \mathrm{dim}\gf\}$ are chosen in such way that $\mathcal{K}(J^{a},J^{b})=h^{\vee}\delta_{ab}$ and $\{J^{a};\; a=\mathrm{dim}\gf-\mathrm{dim}\af\dots \mathrm{dim}\gf\}$ are the generators of subalgebra $\af\subset \gf$. Now we can consider $(\mathrm{dim}\gf-\mathrm{dim}\af)$-dimensional Brownian motion with generators $d\theta^{a}$ such that $\mathbb{E}(d\theta^{a} \; d\theta^{b})=\delta_{ab};\; a,b=1,\dots,\mathrm{dim}\gf-\mathrm{dim}\af$.

Then for generator of field transformation we can write 
\begin{equation}
  \mathcal{G}_{i}=\left(\frac{2dt}{w_{i}}-\sqrt{\kappa} d\xi_{t}\right) \partial_{w_{i}}+\frac{\sqrt{\tau}}{w_{i}}\left(\sum_{a=1}^{\mathrm{dim}\gf-\mathrm{dim}\af}\left(d \theta ^{a} t^{a}_{i}\right)\right)
\label{eq:5}
\end{equation}
Substituting to equation \eqref{eq:8} we get martingale condition

\begin{equation}
  \left(-2 \mathcal{L}_{-2}+\frac{1}{2}\kappa \mathcal{L}_{-1}^{2}+\frac{\tau}{2}\left( \sum_{a=1}^{\mathrm{dim}\gf-\mathrm{dim}\af} \mathcal{J}^{a}_{-1} \mathcal{J}^{a}_{-1}\right)\right)        \mathcal{F}_{\mathbb{H}}=0
\label{eq:9}
\end{equation}

which can be rewritten as the requirement for
\begin{equation}
  \psi=\left(-2L_{-2}+\frac{1}{2}\kappa L_{-1}^{2}+\frac{1}{2}\tau \left(\sum_{a=1}^{\mathrm{dim}\gf-\mathrm{dim}\af}J^{a}_{-1}J^{a}_{-1}\right)\right) \phi_{(\mu,\nu)}
\label{eq:10}
\end{equation}
to be level two null-field. 

Now consider simple example. Let $G=SU(2)$ and $A=U(1)$, and corresponding Lie algebras $\gf=\mathrm{su}(2)$
with generators $J^{1},J^{2},J^{3}$ and $\af=\mathrm{u}(1)$ with the generator $J^{3}$, $\af\subset\gf$. Note that $\mathcal{K}(J^{a},J^{b})=2\delta^{ab}$. 

 The equation \eqref{eq:10} is now
\begin{equation}
  \label{eq:11}
  \psi=\left(-2L_{-2}+\frac{1}{2}\kappa L_{-1}^{2}+\frac{1}{2}\tau \left(J^{1}_{-1}J^{1}_{-1}+J^{2}_{-1}J^{2}_{-1}\right)\right) \phi_{(\mu,\nu)}
\end{equation}
If we use basis $J^{+}=\frac{J^{1}+iJ^{2}}{\sqrt{2}},\; J^{-}=\frac{J^{1}-iJ^{2}}{\sqrt{2}}$ this equation is rewritten in the form
\begin{equation}
 \psi= \left(-2 L_{-2}+\frac{\kappa}{2}L_{-1}^{2}+\frac{\tau}{2}\left[J^{+}_{-1}J^{-}_{-1}+J^{-}_{-1}J^{+}_{-1}\right]\right) \phi_{(\mu,\nu)}
\label{eq:12}
\end{equation}
which is similar to the equations for parafermionic fields introduced in paper \cite{santachiara2008sle}.

Central charge in this case is equal to
\begin{equation}
  \label{eq:14}
  c=\frac{2(k-1)}{k+2}
\end{equation}

Let us study the solutions of the equation \eqref{eq:11}. Act on $\psi$ with $L_{2}$:
\begin{equation}
  \label{eq:13}
  L_{2}\psi=(-8L_{0}+c+3\kappa L_{0}+\tau k)\phi_{(\mu,\nu)}=0,
\end{equation}
and obtain
\begin{equation}
  \label{eq:28}
  (3\kappa-8) h_{(\mu,\nu)}+c+\tau k =0
\end{equation}
Another equation can be obtained by action of $L_{1}^{2}$
\begin{equation}
  \label{eq:15}
  L_{1}^{2}\psi = (12 L_{0} + \kappa(4 L_{0}^{2}+2 L_{0}) +\tau (J_{0}^{1}J_{0}^{1}+J_{0}^{2}J_{0}^{2}))\phi_{(\mu,\nu)}=0
\end{equation}
Since $J_{0}^{1}J_{0}^{1}+J_{0}^{2}J_{0}^{2}=J_{0}^{1}J_{0}^{1}+J_{0}^{2}J_{0}^{2}+J_{0}^{3}J_{0}^{3}-J_{0}^{3}J_{0}^{3}$ we have
\begin{equation}
  \label{eq:17}
  12 h_{(\mu,\nu)}+2\kappa h_{(\mu,\nu)} (2h_{(\mu,\nu)}+1) +  \tau (C_{\mu}-\tilde{C}_{\nu})=0,
\end{equation}
where $C_{\mu}=(\mu,\mu+2\rho)$ is the eigenvalue of quadratic Casimir operator $\sum_{a}t^{a}t^{a}$. For $u(1)$ it is just $\nu^{2}$.
We need to get one more equation since we have three variables $\kappa,\tau,h_{(\mu,\nu)}$. We can use that $\psi$ is singular vector so $L_{1}\psi=0$ and then $J_{1}^{3}L_{1}\psi=0$, so
\begin{equation}
  \label{eq:29}
  \left(-6  + \kappa +  \tau +  2 \kappa h_{(\mu,\nu)}\right) J^{3}_{0} \phi_{(\mu,\nu)}=0
\end{equation}

Turn to the case $k=2, \;c=1/2$ which corresponds to square lattice Ising model. We have three non-equivalent fields numbered by the pairs of $su(2)$ and $u(1)$ weights, which are $\phi_{(0,0)}, \; h_{(0,0)}=0; \quad \phi_{(0,2)}, \; h_{(0,2)}=1/2; \quad \phi_{(1,1)}, \; h_{(1,1)}=1/16$.
The equation \eqref{eq:29} is trivial for $\phi_{(0,0)}$ and $\phi_{(0,2)}$, so the equations \eqref{eq:17}, \eqref{eq:28} are consistent although we do not know how to interpret $\tau$ in context of the Ising model. If we consider the field $\phi_{(1,1)}$ we have no solution for $\kappa,\tau$, so we see that not every primary field corresponds to some boundary condition changing operator. 

Now let us generalize to arbitrary coset model. It is not always convenient to work with generators orthogonal with respect to Killing form $\mathcal{K}$, sometimes it is preferable to use Chevalley or Cartan-Weyl basis. So we change normalization condition of additional $\mathrm{dim}\gf$-dimensional Brownian motion \eqref{eq:5} to $\mathbb{E}\left[d\theta^{a}\; d\theta^{b}\right]=\mathcal{K}(t^{a},t^{b})dt$, and the generator of field transformation \eqref{eq:2} is
\begin{equation*}
  \mathcal{G}_{i}=\left(\frac{2dt}{w_{i}}-\sqrt{\kappa} d\xi_{t}\right) \partial_{w_{i}}+\frac{\sqrt{\tau}}{w_{i}}\left(\sum_{a:\mathcal{K}(t^{a},\tilde{t}^{b})=0}\left(d \theta ^{a} t^{a}_{i}\right)\right).
\end{equation*}
 Martingale condition  \eqref{eq:8} can be written as
\begin{equation*}
  \left(-2 \mathcal{L}_{-2}+\frac{1}{2}\kappa \mathcal{L}_{-1}^{2}+\frac{\tau}{2}\left( \sum_{a} \mathcal{J}^{a}_{-1} \mathcal{J}^{a}_{-1}-
      \sum_{b}\tilde{\mathcal{J}}^{b}_{-1} \tilde{\mathcal{J}}^{b}_{-1}\right)\right)        \mathcal{F}_{\mathbb{H}}=0.
\end{equation*}
This equation can again be rewritten as the requirement for
\begin{equation}
  \psi=\left(-2L_{-2}+\frac{1}{2}\kappa L_{-1}^{2}+\frac{1}{2}\tau \left(\sum_{a=1}^{\mathrm{dim}\gf}J^{a}_{-1}J^{a}_{-1}-\sum_{b=1}^{\mathrm{dim}\af}\tilde{J}^{b}_{-1}\tilde{J}^{b}_{-1}\right)\right) \phi_{(\mu,\nu)}
\label{eq:30}
\end{equation}
to be level two null-field.

Acting on \eqref{eq:30} with $L_{2}$ and $L_{1}^{2}$ we can obtain equations \eqref{eq:28} and \eqref{eq:17}. To get more equations we can act by $\afh$-generators $\tilde{J}^{b}_{1}$ and $L_{1}$ similarly to what was done with $J^{3}_{1}$ to get equation \eqref{eq:29}. 

To obtain simpler relations it is also possible to use generalization of Knizhnik-Zamolodchikov equations \eqref{eq:6} as was done for WZNW-models in paper \cite{alekseev2010sle}. 

\section{Conclusion}
\label{sec:conclusion}

We proposed a way to generalize Schramm-Loewner evolution to obtain observables which can be studied by methods of coset conformal field theory. The description of fields in coset models is not very simple due to field identification \cite{schellekens1990field} and the need of  fixed points resolution \cite{Fuchs:1996dd,fuchs1996resolution} which we have not discussed here. These subtleties can complicate the solution of equations  \eqref{eq:28}, \eqref{eq:17} and analogues of equation \eqref{eq:29}. On the other hand the use of Knizhnik-Zamolodchikov equations \cite{kogan1997knizhnik} for correlation functions in the spirit of \cite{alekseev2010sle} leads to matrix algebraic relations which are similar to NIM-representations for boundary states \cite{ishikawa2003novel}. The study of this subject can reveal deep algebraic connection of martingale conditions with boundary states classification. 

Physical interpretation of the solutions of martingale conditions is not clear, but there is also no lattice interpretation of Schramm-Loewner evolution with additional Brownian motion on group manifold \cite{bettelheim2005stochastic} and no lattice interpretation of WZNW-models. We think that these questions are closely related.

\section*{Acknowledgements}
\label{sec:acknowledgements}
I thank the organizers of The Seventh International Conference ``Quantum Theory and Symmetries'' (QTS-7) for the opportunity to present this work and especially professor C. Burdik for his attention. I am grateful to A. Bytsko, K. Izyurov, D. Chelkak and R. Santachiara for the discussions on SLE and WZNW models. This work  is supported by
the Chebyshev Laboratory (Department of Mathematics and Mechanics,
Saint-Petersburg State University) under the grant 11.G34.31.0026
of the Government of the Russian Federation.

\bibliography{bibliography}{}

\providecommand{\newblock}{}
\begin{thebibliography}{10}
\expandafter\ifx\csname url\endcsname\relax
  \def\url#1{{\tt #1}}\fi
\expandafter\ifx\csname urlprefix\endcsname\relax\def\urlprefix{URL }\fi
\providecommand{\eprint}[2][]{\url{#2}}

\bibitem{schramm2000scaling}
Schramm O 2000 {\em Israel Journal of Mathematics\/} {\bf 118} 221--288

\bibitem{rohde2005basic}
Rohde S and Schramm O 2005 {\em Annals of mathematics\/}  883--924

\bibitem{bauer20062d}
Bauer M and Bernard D 2006 {\em Physics reports\/} {\bf 432} 115--221

\bibitem{Cardy:2005kh}
Cardy J~L 2005 {\em Annals Phys.\/} {\bf 318} 81--118 (\textit{Preprint}
  \eprint{cond-mat/0503313})

\bibitem{bauer2004conformal}
Bauer M and Bernard D 2004 Conformal transformations and the sle partition
  function martingale {\em Annales Henri Poincare\/} vol~5 (Springer) pp
  289--326

\bibitem{bauer2004cfts}
Bauer M and Bernard D 2004 {\em Physics Letters B\/} {\bf 583} 324--330

\bibitem{bauer2003sle}
Bauer M and Bernard D 2003 {\em Physics Letters B\/} {\bf 557} 309--316

\bibitem{bauer2002sle}
Bauer M and Bernard D 2002 {\em Physics Letters B\/} {\bf 543} 135--138

\bibitem{bettelheim2005stochastic}
Bettelheim E, Gruzberg I, Ludwig A and Wiegmann P 2005 {\em Physical review
  letters\/} {\bf 95} 251601

\bibitem{Rasmussen:2004xr}
Rasmussen J 2007 {\em Afr.J.Math.Phys.\/} {\bf 4} 1--9 (\textit{Preprint}
  \eprint{hep-th/0409026})

\bibitem{alekseev2010sle}
Alekseev A, Bytsko A and Izyurov K 2010 {\em Letters in Mathematical Physics\/}
   1--19

\bibitem{santachiara2008sle}
Santachiara R 2008 {\em Nuclear Physics B\/} {\bf 793} 396--424

\bibitem{picco2008numerical}
Picco M and Santachiara R 2008 {\em Physical review letters\/} {\bf 100} 15704

\bibitem{difrancesco1997cft}
Di~Francesco P, Mathieu P and Senechal D 1997 {\em {Conformal field theory}\/}
  (Springer)

\bibitem{fuchs2005geometry}
Fuchs J and Wurtz A 2005 {\em Nuclear Physics B\/} {\bf 724} 503--528

\bibitem{fredenhagen2002d}
Fredenhagen S and Schomerus V 2002 {\em Journal of High Energy Physics\/} {\bf
  2002} 005

\bibitem{elitzur2002d}
Elitzur S and Sarkissian G 2002 {\em Nuclear Physics B\/} {\bf 625} 166--178

\bibitem{Maldacena:2001ky}
Maldacena J~M, Moore G~W and Seiberg N 2001 {\em JHEP\/} {\bf 07} 046
  (\textit{Preprint} \eprint{hep-th/0105038})

\bibitem{felder1999geometry}
Felder G, Fr{\"o}hlich J, Fuchs J and Schweigert C 1999 {\em Arxiv preprint
  hep-th/9909030\/}

\bibitem{alekseev1999d}
Alekseev A and Schomerus V 1999 {\em Physical Review D\/} {\bf 60} 061901

\bibitem{Polyakov:1970xd}
Polyakov A~M 1970 {\em JETP Lett.\/} {\bf 12} 381--383

\bibitem{smirnov2007towards}
Smirnov S 2007 {\em Arxiv preprint arXiv:0708.0032\/}

\bibitem{duminil2011conformal}
Duminil-Copin H and Smirnov S 2011 {\em Arxiv preprint arXiv:1109.1549\/}

\bibitem{bauer2003conformal}
Bauer M and Bernard D 2003 {\em Communications in mathematical physics\/} {\bf
  239} 493--521

\bibitem{cardy1984conformal}
Cardy J 1984 {\em Nuclear Physics B\/} {\bf 240} 514--532

\bibitem{cardy1989boundary}
Cardy J 1989 {\em Nuclear Physics B\/} {\bf 324} 581--596

\bibitem{cardy1991bulk}
Cardy J and Lewellen D 1991 {\em Physics Letters B\/} {\bf 259} 274--278

\bibitem{belavin1984ics}
Belavin A, Polyakov A and Zamolodchikov A 1984 {\em Nuclear Physics\/} {\bf
  241} 333--380

\bibitem{Goddard198588}
Goddard P, Kent A and Olive D 1985 {\em Physics Letters B\/} {\bf 152} 88 -- 92
  ISSN 0370-2693

\bibitem{gawdzki1988g}
Gawdzki A {\em et~al.\/} 1988 {\em Physics Letters B\/} {\bf 215} 119--123

\bibitem{fuchs1996resolution}
Fuchs J, Schellekens B and Schweigert C 1996 {\em Nuclear Physics B\/} {\bf
  461} 371--404

\bibitem{schellekens1990field}
Schellekens A and Yankielowicz S 1990 {\em Nuclear Physics B\/} {\bf 334}
  67--102

\bibitem{kogan1997knizhnik}
Kogan I, Lewis A and Soloviev O 1997 {\em Arxiv preprint hep-th/9703028\/}

\bibitem{Fuchs:1996dd}
Fuchs J, Schellekens A~N and Schweigert C 1996 {\em Nucl. Phys.\/} {\bf B473}
  323--366 (\textit{Preprint} \eprint{hep-th/9601078})
  \urlprefix\url{http://www.nikhef.nl/~t58/kac.html}

\bibitem{ishikawa2003novel}
Ishikawa H and Tani T 2003 {\em Nuclear Physics B\/} {\bf 649} 205--242

\end{thebibliography}
\bibliographystyle{iopart-num}

\end{document}